# Interior and edge magnetization in thin exfoliated CrGeTe$_3$ films

*Avia Noah\*[1], Hen Alpern[1,2]\*\*, Sourabh Singh[1], Alon Gutfreund[1], Gilad Zisman[1], Tomer D. Feld[1], Atzmon Vakahi[3], Sergei Remennik[3], Yossi Paltiel[2], Martin E. Huber[4], Victor Barrena[5], Hermann Suderow[5], Hadar Steinberg[1], Oded Millo[1], and Yonathan Anahory\*\*\*[1]*

[1]The Racah Institute of Physics, The Hebrew University, Jerusalem, 91904, Israel
[2]Department of Applied Physics, The Hebrew University of Jerusalem, Jerusalem 91904, Israel
[3]Center for Nanoscience and Nanotechnology, Hebrew University of Jerusalem, Jerusalem, 91904, Israel
[4]Departments of Physics and Electrical Engineering, University of Colorado Denver, Denver, CO 80217, USA
[5]Laboratorio de Bajas Temperaturas, Unidad Asociada UAM/CSIC, Departamento de Física de la Materia Condensada, Instituto Nicolás Cabrera and Condensed Matter Physics Center, Universidad Autónoma de Madrid, E-28049 Madrid, Spain

KEYWORDS: 2D magnetism, edge magnetism, superconducting/ferromagnetic interface, scanning SQUID-on-tip microscopy, CrGeTe$_3$

Corresponding authors: *avia.noah@mail.huji.ac.il, **alpernhen@gmail.com,
*** yonathan.anahory@mail.huji.ac.il

## Abstract

CrGeTe$_3$ (CGT) is a semiconducting vdW ferromagnet shown to possess magnetism down to a two-layer thick sample. Although CGT is one of the leading candidates for spintronics devices, a comprehensive analysis of CGT thickness dependent magnetization is currently lacking. In this work, we employ scanning SQUID-on-tip (SOT) microscopy to resolve the magnetic properties of exfoliated CGT flakes at 4.2 K. Combining transport measurements of CGT/NbSe2 samples with SOT images, we present the magnetic texture and hysteretic magnetism of CGT, thereby matching the global behavior of CGT to the domain structure extracted from local SOT magnetic imaging. Using this method, we provide a thickness dependent magnetization state diagram of bare CGT films. No zero-field magnetic memory was found for films thicker than 10 nm and hard ferromagnetism was found below that critical thickness. Using scanning SOT microscopy, we identify a unique edge magnetism, contrasting the results attained in the CGT interior.

## Introduction

Layered van der Waals (vdW) ferromagnets have recently been the focus of intensive research due the easily accessible broad thickness range they offer, from the bulk material all the way to atomically-thin two-dimensional (2D) crystals, enabled by exfoliation. While the revolution triggered by the vdW materials is well underway [1–4], the emerging field of 2D vdW spintronics is still at its infancy[5–7]. The need for compatible materials with long-range ferromagnetic order and precise analysis of such materials are at the core of this new emerging field. The evolution of the magnetic properties from bulk material to thin exfoliated layers may offer additional insight into the origin of ferromagnetism in vdW materials, where anisotropy was suggested[8] to originate from distinct inter-layer and intra-layer exchange interactions. Exfoliating bulk vdW ferromagnets, either conducting such as Fe$_3$GeTe$_2$ (FGT)[9], or semiconducting CrGeTe$_3$

(CGT)[8] and CrI$_3$[10] has revealed that ferromagnetism can survive down to the few layers regime where the Mermin-Wagner theorem asserts long-range ordering should be suppressed by thermal fluctuations in the absence of magnetic anisotropy[11]. Such anisotropy is manifested as out-of-plane (OOP) easy axis magnetization for both FGT[12] and CGT[13,14]. Ferromagnetism in those materials was mostly characterized using Anomalous Hall effect (AHE) measurements (that cannot be applied to the insulating CGT)[9,15–17] and SQUID (superconducting quantum interference device) magnetometry[12,13,18], which average over the whole sample, or by local probes such as Kerr rotation[8], low temperature magnetic force microscopy (MFM)[12], Xray magnetic circular dichroism (XMCD)[18] and NV-Centers[19]. The local probe methods are highly effective for investigating edge magnetization in vdW materials, an issue that has recently attracted considerable interest.

In our present work we utilize scanning SQUID-on-tip (SOT), with high spatial resolution [20–22] and single-electron magnetic moment sensitivity[23,24], in combination with transport measurements of CGT/NbSe$_2$ bilayers, to provide an accurate thickness dependence of the magnetic properties of CGT flakes. Our results show that the magnetic characteristics at the flake's edges is different from its interior. The thickness dependence of the film's magnetic behavior can offer a control mechanism that could be used in GMR like devices.

**Results**

*CGT/NbSe$_2$ bilayer*

Probing the magnetic properties of ferromagnetic materials using electrical measurements such as AHE is a powerful method to study samples that are too small to be characterized by bulk magnetization techniques. However, insulating materials such as CGT are not compatible with electrical measurements. Hence, so far the magnetism of CGT was characterized only indirectly by performing transport measurements on a conducting layer coupled to CGT, including induced AHE in proximitized Pt[16], topological insulator (TI) [17] and through magnetoresistance hysteresis in a ferromagnet/superconductor CGT/NbSe$_2$ bilayer[25].

The CGT/NbSe$_2$ sample presented in Fig. 1 consists of ~30nm CGT flake placed on a ~30nm NbSe$_2$ exfoliated on top of pre-patterned Au contacts (see Fig. S6). Fig. 1a presents the longitudinal resistance ($R_{xx}$) of the NbSe$_2$ flake with constant current $I_x = 250$ μA as a function of the out-of-plane (OOP) magnetic field $\mu_0 H_z$ at 4.2 K. In this magnetoresistance measurement, $\mu_0 H_z$ was ramped up from 0 to 130 mT (blue curve) and ramped down back to 0 (red curve). A clear hysteresis is evident between $\mu_0 H_z = 40$ mT and $\mu_0 H_z = 80$ mT, where a switching between the dissipationless and resistive states occurs, consistent with previous measurements reported in Ref. [25], yet its origin was not explained.

To gain better insight into the origin of this hysteretic behavior, we conduct local magnetic field imaging $B_z(x,y)$ using a scanning SOT, aiming to correlate the local magnetic structure of the CGT flake and the magnetoresistance hysteresis of the bilayer. The SOT measurements were simultaneously performed with the transport using SOT with loop diameter ranging from 155 to 180 nm (see Methods and Supplementary Note 1). Figure 1b presents a SOT image of the CGT sample measured at a distance of ~100 nm above the sample for $\mu_0 H_z = 0$. The image resolves magnetic domain features sized lower than the tip diameter (155 nm), yielding a magnetic contrast of ~1 mT. With increasing OOP field, domains parallel to the field grow at the expense of the anti-parallel domains (Fig. 1b-d and Supplementary movie 1). Above the saturation field, $\mu_0 H_S \sim 100$ mT, the magnetic landscape becomes smooth, with a weaker contrast. These results are consistent with the transport and global magnetization measurements of Pt/CGT(65 nm) bilayers[16] as well as with the general behavior of bulk CGT [13,26]. By decreasing the field, the sample's magnetic images remain featureless down to $\mu_0 H_z = 40$ mT (Fig. 1d,e and Fig. 1a right inset), where magnetic domains re-appear (Fig. 1f and Fig. 1a left inset).

A clear correlation emerges between the transport measurement and the magnetic images. The magnetic texture of the CGT flake (Fig. 1b) is expected to provide local pinning potentials, inhibiting flux flow, which is manifested as the zero-voltage state (Fig. 1a blue curve prior to point c). Upon saturating the CGT magnetization (Fig. 1c), the pinning potential flattens, allowing flux flow that generates dissipation and hence a finite voltage. Once the CGT is fully magnetized (Fig. 1d, right inset), the pinning potential is sufficiently uniform to yield un-inhibited flux flow manifested in a linear magnetoresistance[27]. When reducing the field back from the saturation field, the linear magnetoresistance persists (Fig. 1a red curve), in agreement with the featureless images (Fig. 1d,e). An abrupt formation of magnetic domains takes place at a demagnetization field, $\mu_0 H_d = 40$ mT. Importantly, the CGT's demagnetization (Fig. 1f) occurs simultaneously with the switching of the transport measurements back to the dissipationless state where vortices are pinned by the magnetic structure (Fig. 1 left inset).

Our SOT images thus provide a clear evidence for the magnetic texture of the CGT causing the hysteretic magnetoresistance observed on the CGT/NbSe$_2$ bilayer (Fig. 1a). Furthermore, due to the exact correlation between the transport measurements and the magnetic imaging, we demonstrate how the magnetoresistance of the CGT/NbSe$_2$ bilayer could be used to globally probe the magnetic properties of the CGT flake.

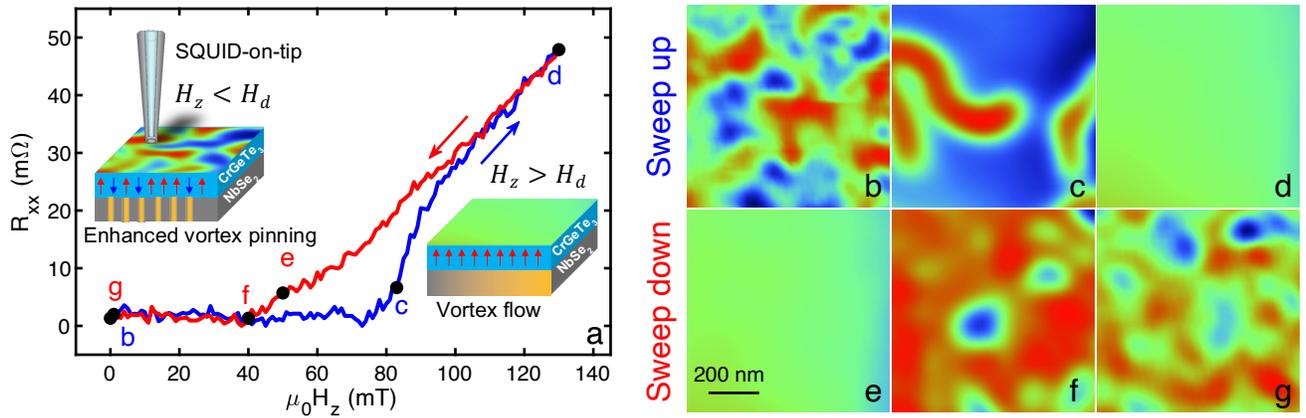

**Fig. 1 CrGeTe$_3$/NbSe$_2$ magnetoresistance and corresponding SOT images at 4.2K. (a)** $R_{xx}$ measurements of the NbSe$_2$ layer as a function of out-of-plane (OOP) magnetic field $\mu_0 H_z$ while applying current I$_x$ = 250 μA and sweeping the field up/down (blue/red) **Left inset**: Schematic illustration of the bilayer below the demagnetization field $H_d = 40$ mT in the enhanced vortex pinning state. Stationary vortices are depicted in orange **Right inset:** Same as left but above $H_d$ in the vortex flow (finite $R_{xx}$) state. **(b-g)** Sequence of magnetic images of the OOP component of the local magnetic field $B_z(x,y)$ of different states of the CGT at distinct values of $\mu_0 H_z$ acquired simultaneously with transport data in **a** (labeled black dots). **(b-d)** $B_z(x,y)$ images acquired sweeping up the field at $\mu_0 H_z = 0$ **b**, 85 **c**, 130 **d** mT. **(e-g)** $B_z(x,y)$ images acquired sweeping the field down $\mu_0 H_z = 50$ **e**, 40 **f**, 0 **g** mT. All images are 1x1 μm² in size, pixel size 20 nm, acquisition time 5 min/image. The blue to red color scale represents lower and higher magnetic field, respectively, with a shared scale of $B_z = 1$ (**b,g**), 5 (**c-f**) mT. See Supplementary Movie 1.

It is worth noting that both the magnetic images and the transport measurement indicate magnetic hysteresis between $\mu_0 H_z = 40$ mT and $\mu_0 H_z = 80$ mT and the CGT demagnetizes at a positive field. Fig. 1b and Fig. 1g show very similar domain structure at zero field both before and after the saturation field $H_s$ was attained. However, the magnetic images alone cannot provide a definitive answer as to whether CGT holds any magnetization at zero field, or does the CGT lose any magnetic memory in the absence of applied field. To describe the magnetic behavior near zero field, we saturated the sample by applying large opposite fields, $\mu_0 H_z = \pm 1T$, before ramping the field back to zero and performed transport measurements and magnetic imaging between $\mu_0 H_z = 0$ mT to $\mu_0 H_z = 130$ mT (See Fig. 2a). By employing this protocol, any memory that the CGT might hold at zero field will be manifested as deviations in the magnetoresistance and magnetic imaging between the two excursions at either $\mu_0 H_z = +1$ T or $\mu_0 H_z = -1$ T. The two magnetoresistance curves presented in Fig. 2a, taken after negative/positive excursions (blue/red curves) show no measurable difference between them. The magnetic images also appear to be insensitive to the change in initial conditions. Figure 2c-f show the same type of features as a function of the field as figure 2g-j (movie 2). Both local (images) and global (transport) measurements show no measurable memory effect for ~30 nm thick CGT.

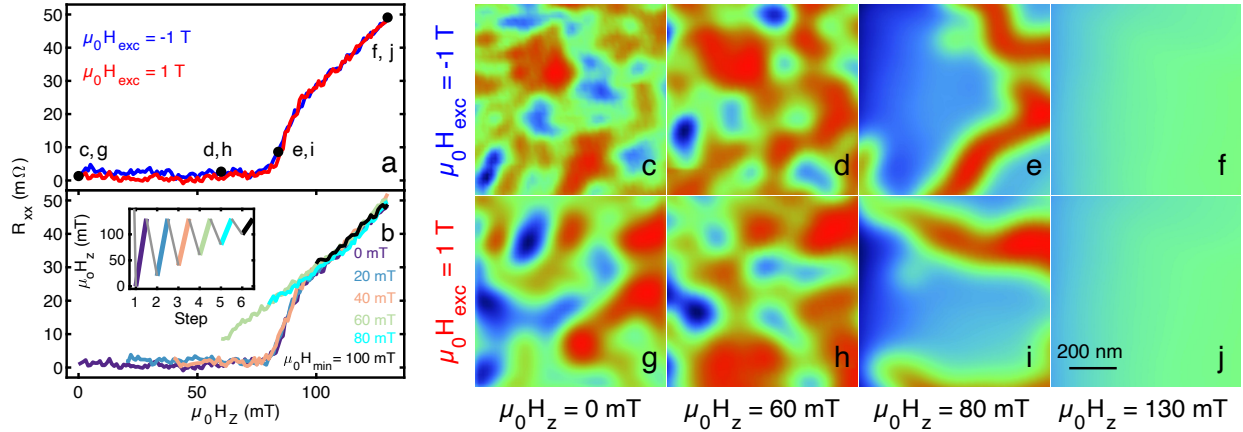

**Fig. 2 SOT images and transport of bilayer CrGeTe$_3$/NbSe$_2$ showing no magnetic memory at $\mu_0 H_z = 0$ at 4.2K.** (a) Evolution of $R_{xx}$ as a function of the out-of-plane (OOP) field $H_z$ from zero to above the saturation field $H_s$ after an excursion to $\mu_0 H_{ex} = -1$ T (blue) and $\mu_0 H_{ex} = 1$ T (red). (b) Evolution of $R_{xx}$ from $\mu_0 H_z = \mu_0 H_{min}$ to $\mu_0 H_z > \mu_0 H_s$ after applying OOP field $\mu_0 H_{max} = 130$ mT saturating the sample. $\mu_0 H_{min} = 0, 20, 40, 60, 80, 100$ mT. **Inset** History of $H_z$ during $R_{xx}$ measurements shown in **b** with color corresponding to the segment color. Grey segments are not shown. (**c-j**) Sequence of magnetic images of different states of the CGT at distinct $H_z$ values. Evolution from $\mu_0 H_z = 0\ mT$ to $\mu_0 H_z > \mu_0 H_s$ after $H_{ex} = -1$ T **c-f** and $H_{ex} = 1$ T **g-j**. $\mu_0 H_z = 0$ **c,g**, 60 **d,h**, 80 **e,i**, 130 mT **f,j** All images are 1×1 μm², pixel size is 20 nm and acquisition time 5 min/image. The blue to red color scale represents lower and higher magnetic field, respectively, with a shared scale of $B_z = 1$ (**c,g**), 5 (**d-f,h-j**) mT. See Supplementary Movies 2.

The measurements shown in Fig. 1 show that the ~30 nm CGT flake retains magnetic memory, and therefore is hysteretic, only in the $\mu_0 H_z = 40 - 80$ mT field range. To verify that the sample loses memory at higher fields than zero, we ramped down the field between increasing minimal fields $\mu_0 H_{min} = 0, 20, 40, 60, 80$, and 100 mT while keeping the maximum field constant and above the saturation field $\mu_0 H_{max} = 130$ mT. An illustration of the measurement scheme is presented in the inset of Fig. 2b. By not ramping down the field to zero, it is expected that more domains pointing with the field will act as nucleation centers to change the field at which the sample is fully magnetized[28,29]. The magnetoresistance curves are shown in Fig. 2b. The transport measurements reveal that the CGT is hysteretic only when $\mu_0 H_{min} > 40$ mT, i.e., the CGT show no measurable memory effect below $\mu_0 H_z = 40$ mT, in excellent agreement with magnetic images that indicate 40 mT to be the demagnetization field.

The data presented in Fig. 1 and Fig. 2 show two key points: that ~30 nm CGT does not hold any long range magnetism below $H_d$, and that the CGT flake globally demagnetizes abruptly at a field indicated by the local magnetic images (Fig. 1f). Importantly, the magnetic images lend themselves to determine $H_s$ and $H_d$ even without the need of NbSe$_2$ (or any other) metallic layer, as shown in the following.

*Thickness dependence of CGT magnetization*

We now turn to the thickness dependence of the saturation and demagnetization fields. We use the SOT to image areas of distinct thickness $d$ on various CGT flakes (Fig, 3a-l). For areas where $d \gtrsim 10$ nm, the magnetic images presented in Fig. 3a-h are used to find the values of $H_s$ and $H_d$. These values are then plotted in Figure 3m and connected to each other with a dashed line giving rise to a bowtie hysteresis loop (Figure 3m, top two sketched curves). Thinner films yield lower values of $H_d$ and $H_s$. For $d \lesssim 10$ nm, the two hysteretic parts of the loop merge and the sample behaves like a standard ferromagnet with an open hysteresis loop (Figure 3m, bottom curve). This is seen in figure 3i and 3l where the sample stays fully magnetized at zero field, in contrast with thicker area of the flake where the sample demagnetizes

(Fig 3 a,d,e,h). A comprehensive thickness-dependent of sketched magnetization curves for a broad range of CGT thicknesses is plotted in Fig. S3. Transport measurements similar to those shown in figure 1 and 2 were performed for a $d < 10$ nm CGT flake manifesting zero-field magnetization effect (See supplementary Figure S11 and Note 4).

In figure 3n, we summarize the values of $H_d$ (green dots) and $H_s$ (red dots) for all the imaged thicknesses. The lines connecting these points constitute borders between distinct magnetic states; the domains state (purple), the hysteretic state (orange) and the fully magnetized state (blue). In the domains state, the CGT exhibits small magnetic domains that are insensitive to the excursion field, whereas the opposite holds for the fully magnetized region. In the hysteretic region, the sample can be either in the fully magnetized state or in the domains state depending on the applied magnetic field history.

The thickness dependence of CGT magnetization was measured here for pristine exfoliated single crystals. The recorded critical thickness for holding magnetization in zero field, ~10 nm, is seemingly not in agreement with a few other AHE works conducted on CGT [16,17,30], where thicker layers of CGT seem to attain magnetic memory at zero field (finite $R_{xx}$ at $\mu_0 H_z = 0$). This might be because the above works all considered CGT proximitized to large spin orbit materials such as Pt [16] or topological insulators (TIs) such as $Bi_2Te_3$ [17] or $(Bi,Sb)_2Te_3$ [30]. Enhanced magnetism due to hybridization of an insulating ferromagnet to a TI was also seen in a EuS/TI bilayer [31]. Moreover, magnetic anisotropy is heavily generated due to the material spin orbit and hence modifications of that property through proximity can adjust the magnitude of the magnetic anisotropy which, in turn, alters the magnetic properties of the ferromagnet interface [32]. We did not observe any influence on the magnetic structure due to the presence of $NbSe_2$ probably because of a small spatial gap at the interface (Fig. S4) hindering such a proximity effect.

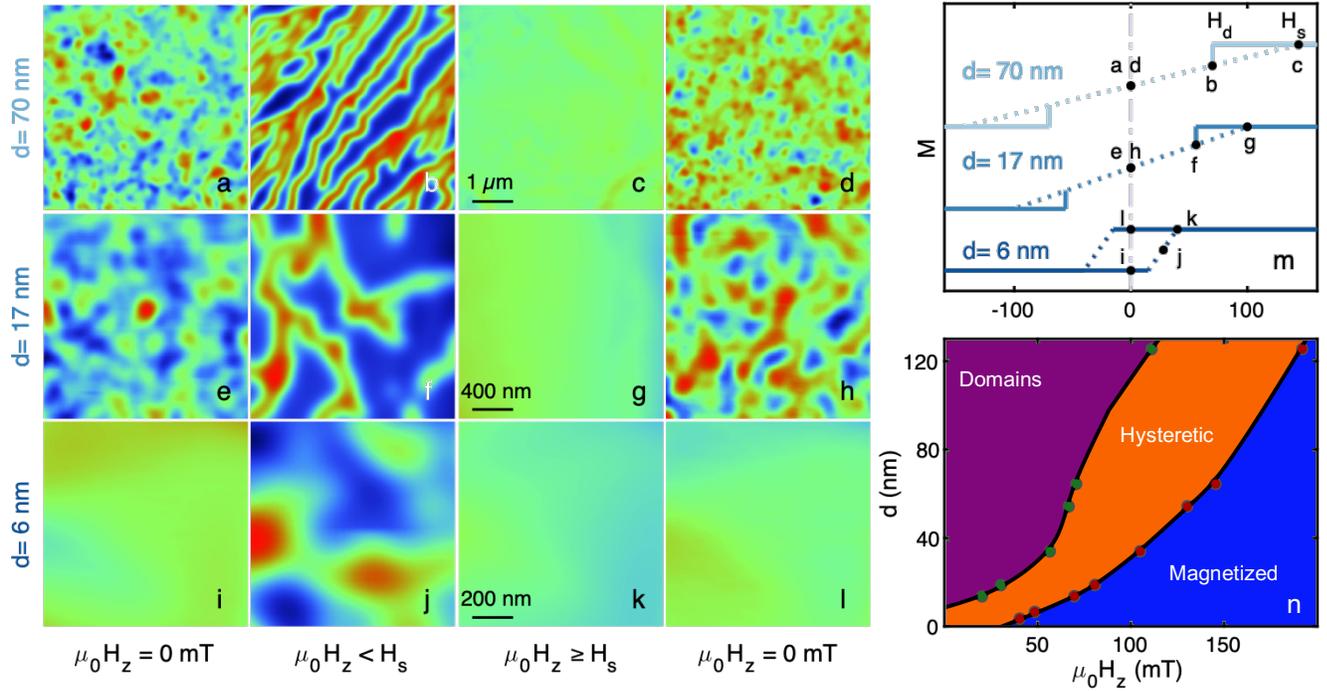

**Fig. 3 Scanning SOT microscopy images of CrGeTe₃ for different thickness at 4.2 K.** (**a-l**) Sequence of magnetic images $B_z(x,y)$ of different states of the sample at distinct values of applied out-of-plane (OOP) field $\mu_0 H_z$ and sample thickness $d$. (**m**) Sketched magnetization curves drawn from $B_z(x,y)$ measured on film's parts of different $d$. Dashed lines are a guide to the eye connecting the two saturated fields. $d = 70$ nm (light blue), $d = 17$ nm (blue), $d = 6$ nm (dark blue). The fields at which the images were taken are marked with black dots. (**n**) A thickness-dependent magnetization-state diagram of CGT showing three states: domains (purple), hysteretic (orange), and magnetized (blue), Imaging parameters: (**a-d**) $d = 70$ nm, area scan 5×5 μm², pixel size 40 nm. $\mu_0 H_z = 0$ **a**, 115 **b**, 175 **c**, 0 mT **d**. (**e-h**) $d = 17$ nm, area scan 2×2 μm², pixel size 30 nm. $\mu_0 H_z = 0$ **e**, 70 **f**, 120 **g**, 0 mT **h**. (**i-l**) $d = 6$ nm, area scan 1x1 μm², pixel size 30 nm. $\mu_0 H_z = 0$ **i**, 20 **j**, 120 **k**, 0 mT **l**. The blue to red color scale represents lower and higher magnetic field, respectively, with a shared scale of $B_z = 1$ (**a,c-e,h-l**), 5 (**b,f,g**) mT. See Supplementary Movies 3-5 corresponding to images **c-f**, **g-j**, **k-n**, respectively. The scale bars in **c**, **g** and **k** apply to all images in the respective row. The x-axis labels of **m** and **n** are the same.

*Edge magnetization*

Another possible explanation for the difference between ours and previous results is that stronger magnetism is concentrated in small regions of the sample. These ferromagnetic regions might have been overrepresented in the AHE measurements performed by other groups. With that potential contradiction in mind, we carefully imaged distinct areas of the sample. We discovered that for thick regions that show a bowtie hysteresis loop, i.e., when the flake interior breaks into domains at $H_d$, its edge retains a magnetic memory. In Fig. 4, we present two sets of images measured at $\mu_0 H_z = 0$ mT after OOP field excursion to $|H_{ex}| > H_s^{\pm}$. In these conditions, domains appear in the CGT interior, but the edge clearly holds the previous magnetization direction (negative or positive - blue or red in Fig. 4), determined by the polarity of previous excursion $H_{ex}$, showing only small fluctuations in $B_z(x,y)$. The flake thicknesses presented in Fig. 4 are 17 nm (Fig. 4a,c) and 24 nm (Fig. 4b,d). The excursion fields magnetizing the sample were: $H_{ex} = \pm 1$ T (Fig. 4a,c) and $H_{ex} = 200$ mT (Fig 4b,d). For samples below the critical thickness, both edge and interior behave like a hard ferromagnet and no edge magnetization is visible (Figure S8).

To try to elucidate this surprising effect, we acquired cross-sectional scanning transmission electron microscopy (STEM) images seen in Fig. 4e,f. The images reveal both the exact thickness of the measured CGT flakes, and the roughness of the edge. Importantly, the edge of the sample has a tapered cross-section, thinning over a lateral distance of 10-20 nm. The average $B_z(x,y)$ calculated along lines in the vertical (y) direction as a function of x position in Figs. 4c and 4d are presented as blue lines in Fig 4g and 4h, respectively. The average $B_z(x)$ signal peaks at ~0.55 mT and ~0.38 mT (and similar values are found for opposite excursion fields) while the inner region remains below 0.25 mT.

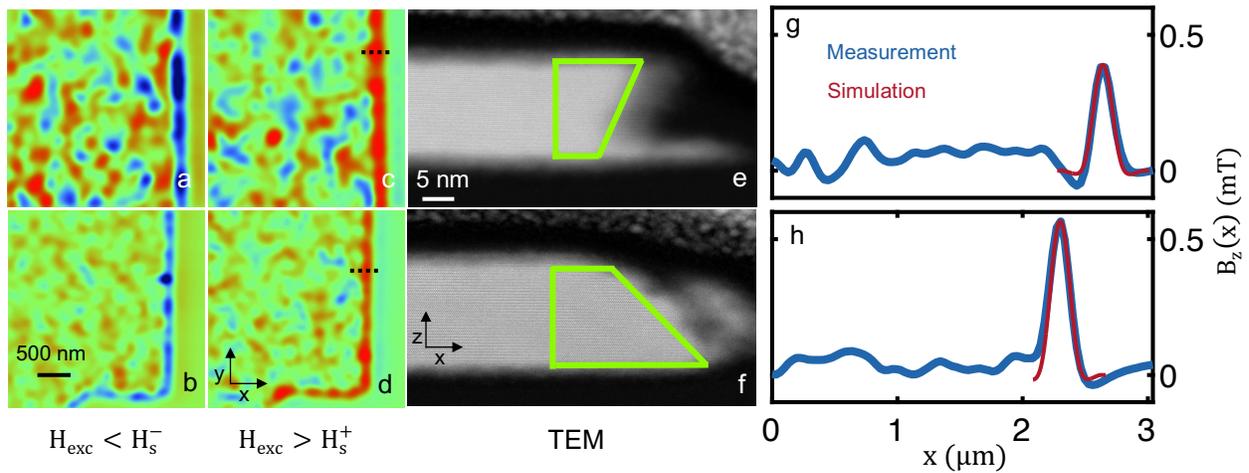

**Fig. 4 Scanning SOT microscopy images of CrGeTe₃ flake interior and edge at zero field and STEM images of the edge.** (**a-d**) Sequence of magnetic images acquired on two different regions of the same flake after distinct field excursions. (**a,b**) $H_{exc} < H_s^-$, (**c,d**) $H_{exc} > H_s^+$. The areas thicknesses are $d = 17$ (**a,c**), 24 nm (**b,d**). (**e,f**) STEM cross-sectional images measured on the black lines presented in **c** and **d**, respectively, lines are not to scale. (**g,h**) Blue lines - an average of the local magnetic field $B_z(x,y)$ along the vertical (y) direction of panels c and d, respectively. Red dashed lines represent the simulations of the edge magnetization stemming from magnetized edges with a trapezoid cross section, marked by green lines in panels e and f, respectively. Imaging parameters: $\mu_0 H_z = 0$ mT, area scan 3×3 μm², pixel size 31 (**a,c**), 24 (**b,d**) nm. The blue to red color scale represents lower and higher magnetic field, respectively, with a shared scale for of $B_z = 1$ mT.

Discussion:

Our work shows that with decreasing thickness, the saturation field $H_s$ diminishes as well as the demagnetization field $H_d$. This trend persists down to ~10 nm, where for thinner flakes $H_d$ crosses zero, thus enabling the CGT to retain magnetic memory at zero field (Fig. 5a and b). We note that the values of $H_d$ and $H_s$ were observed consistently in different areas of the same thickness irrespective of their lateral dimensions that ranged from a few microns to a few tens of microns. Finally, we also observe hard magnetism at the edges for samples above 10 nm (Fig. 5b).

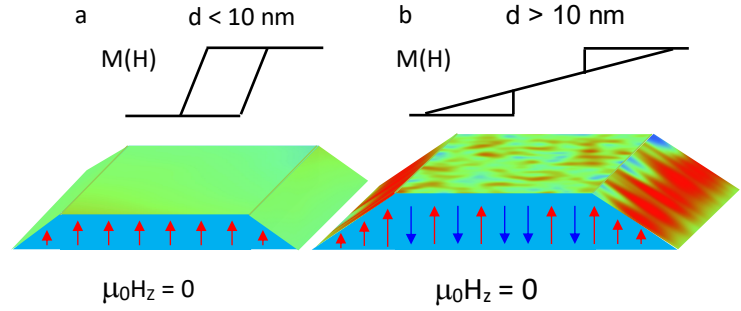

**Figure 5 – Illustration of the local magnetic structure at the edges and in the interior of CrGeTe$_3$ flakes for different thicknesses (a)** Top panel, sketch of the magnetization loop. Bottom panel, local magnetic structure and the resulting out-of-plane magnetic image $B_z(x,y)$ at zero applied field for samples thinner than 10 nm. (**b**) Same as **a** but for sample thicker than 10 nm. The edges retain their magnetization unlike the sample's interior. The magnetization direction of the edge at zero field depends on the field history.

The vanishing remnant magnetization in zero field with increasing thickness is a phenomenon common to a number of vdW ferromagnetic materials[33,34]. The Hamiltonian describing thin OOP magnetized ferromagnets can be written as follows[35]:

$$H = Jf(\vec{x}) - \lambda g(\vec{x}) + \Omega h(\vec{x}), \qquad (1)$$

where $J$ is the exchange integral, $\lambda$ is the effective magnetic anisotropy, $\Omega$ is the strength of the dipole interaction and $f(\vec{x})$, $g(\vec{x})$ and $h(\vec{x})$ are spatial functions of the magnetization. While $J$ and $\lambda$ correspond to local interactions stabilizing the spin magnetization, $\Omega$ is the long-range dipole interaction, making the single domain formation unstable with respect to the creation of stripe domain. Interestingly, when zero field cooling thick CGT flakes, the stripe magnetization is observed (Fig. S7), in agreement with the theoretical prediction in the limit where the dipolar interaction exceeds the magnetic anisotropy[35]. In the case of strong magnetic anisotropy $\lambda$ or larger exchange interaction $J$ the stripes width increases exponentially with these values[36], initiating an approach to the single domain phase. An accurate calculation of $J$, $\lambda$ and $\Omega$ as a function of CGT thickness was not performed to date, though *ab initio* calculations of $J$ and $\lambda$ have shown qualitative agreement with experiments and were seen to change from the 2D to the bulk limit[37]. $J$, $\lambda$ and $\Omega$ are predicted to scale differently as a function of thickness[38], thus inducing a transition from the fragmented domain formation in the thick limit to the hard ferromagnetism in the thin limit. A similar transition was seen for FGT[34] and was accounted for by the same model[18].

We now discuss the edge magnetism (Fig 4 and 5b). The STEM images in Fig. 4e,f reveal a variation of the flake structure on the edge, where its thickness is substantially diminished. Due to the reduced dimensionality of the edge, it is reasonable to postulate that the thinner edge behaves as the thin CGT flake (<10 nm), thereby possessing finite magnetization at zero field (Fig. 5b). Based on this conjecture, we performed magnetostatic simulation of the field profile generated by the thin end of the flake, depicted as a right-angled triangle cross section of area 15x12 nm$^2$. A saturation magnetization of 3 $\mu_B$/Cr with a unit cell volume of 0.83 nm$^3$ [39] was assumed[26]. A convolution of the tip size with the generated stray field at the minimal possible working distance of the SOT (~10 nm) generated an average field of 0.15-0.2 mT, smaller than the 0.38-0.55 mT measured on the edge, yet having the same direction. To better fit the measured data, the saturated section of the flake edge was increased to include a section of the thicker part of the flake as well as the thin edge, constituting trapezoid cross-sections shown in Fig. 4e,f. The simulated magnetism then fits well with the measured data, as can be seen by the dashed red lines in Fig. 4g,h. The simulation fitting yielded a distance of ~100 nm between the SOT and the CGT surface, as expected. Thus, the simulation shows that the edge magnetism has a width of a few tens of nanometers. Fluctuations observed in $B_z(x,y)$ may be due local variations in the effective film thickness, owing to deformations associated with the edge roughness.

The magnetization at the edge could also be explained also by other mechanisms, related to the in-plane dangling bounds. If such mechanisms would be dominant, one should find magnetism also at step-edges between two terraces above the critical thickness. The absence of magnetism at such step-edges (see Figure S9) suggests that this scenario is less probable. We also did not find any preferential oxidation at the flake edge which could account for magnetization there (Fig. S10). The mechanism we propose above thus appears to be plausible one, although others could also be considered.

In conclusion, the presented study demonstrates a direct relation between the global magnetization reading of the CGT by the NbSe$_2$, and local domain structure. The control of the small size domain structure can be utilized to generate highly packed magnetic memory that can be probed by GMR or superconducting wires. Small changes in thickness and edge effects can enhance the memory complexity and external field tuning ability. This effect can be also used in a double layered device with different thicknesses of CGT, where the thick layer will act as the soft magnet and the thinner layer as the hard magnet, which may be useful for spintronics applications.

**TOC Graphic**

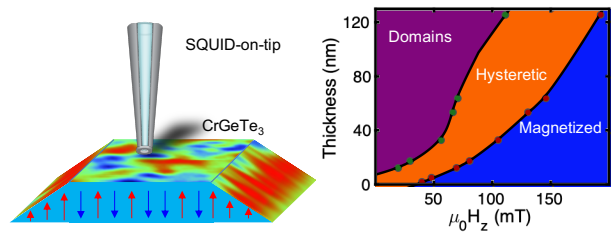

**Methods**

**Sample fabrication**

Bulk NbSe$_2$ was purchased from Graphene HQ.

We grew CGT crystals using the flux method[40,41]. We introduced a mixture of Cr (99.99%), Ge (99.999%) and Te (99.999%) from Goodfellow in a ratio of 1:1:8 in a Canfield crucible set[42,43] and sealed it in a quartz ampoule in an argon atmosphere. We heated to 1000 °C in 12 hours and slowly cooled to 500 °C in 4 days. We removed the ampoule from the furnace and rapidly spinned the crystals to separate the CGT crystals from excess flux. We extracted large crystals, whose size was limited by the size of the crucible. The crystals have shiny surfaces and are plate like. X-ray scattering, magnetization and resistance vs temperature measurements will be published elsewhere and are very similar to previous reports[26,39].

CGT and CGT/NbSe$_2$ bilayer samples were fabricated using the dry transfer technique[44], carried out in a glove-box (argon atmosphere). NbSe$_2$ and CGT flakes were cleaved using the scotch tape method, exfoliated on commercially available Gelfilm from Gelpack. For the transport measurements a NbSe$_2$ flake was transferred onto pre-patterned 50 nm thick Au electrodes fabricated using photolithography on a SiO$_2$ substrate, and a CGT flake was subsequently transferred onto it; both flakes were ~30 nm thick as determined by atomic force microscopy measurements (Figures S2). The samples didn't undergo heating or treatment in any solvents, deeming them pristine (other than naturally occurring oxidation upon removing the samples from the glovebox (see Supplementary Note 3 and Figures S4 and S5).

**Transport measurements.** Transport measurements were performed at 4.2 K inside a Liquid Helium Dewar employing standard 4-probe configuration, where the distances between the current (voltage) contacts were 15 μm (5 μm). Unless otherwise mentioned, a current bias of 250 μA was applied along the ab plane. A magnet consists of a standard SC coil was used to apply out-of-plane (OOP) magnetic fields up to ±1 T.

**Scanning SQUID-On-Tip microscopy:** The SOT was fabricated using self-aligned three-step thermal deposition of Pb at cryogenic temperatures, as described in ref. [23]. Supplementary Figure S1 shows the measured quantum interference pattern of one of the SOTs used for this work with an effective diameter of 155 nm and a maximum critical current of 105 μA. The asymmetric structure of the SOT gives rise to a slight shift of the interference pattern resulting a good sensitivity in zero field. All measurements were performed at 4.2 K in a low pressure He of 1 to 10 mbar.

**Supporting Information**

Additional experimental details and discussion such as; SOT images at ZFC and edge of 6 nm CGT flake, magneto-transport measurements of CGT/NbSe2 with $d < 10$ nm, movies of protocols present in the manuscript, optical, atomic force microscopy, STEM and EDX measurements of the main CGT flake used for the thickness dependence measurements, and information on the SOT parameters.

**Acknowledgements**

We would like to thank S. Gazit, M. Khodas, for fruitful discussions. We would like to acknowledge R. Pradheesh and S. R. K. Chaitanya Indukuri for assistance in sample fabrication and characterization. This work was supported by the European Research Council (ERC) Foundation grant No. 802952 and the Israel Science Foundation (ISF) grant No. 649/17 and 2178/17. The international collaboration on this work was fostered by the EU-COST Action CA16218. H. Suderow thanks J.L. Martínez Peña for informing about the system CrGeTe$_3$ and helping in the initial characterization of the samples, and acknowledges support from Spanish State Research Agency (PID2020-114071RB-I00, CEX2018-000805-M) and the Comunidad de Madrid through program NANOMAGCOST-CM (Program No.S2018/NMT-4321).

**Author contributions**

Y.A., A.N., H.A, O.M., and H. Suderow, conceived the experiment.

A.N. analyzed the data.

A.N. and G.Z. performed the scanning SOT measurements and the transport measurements.

H.A. computed the simulation.

Y.A., S.S., H.A., H. Steinberg and A.N. fabricated and characterized the CGT and CGT/NbSe2 devices.

V.B., and H. Suderow synthesized the CGT crystals.

Y.A., A.N., A.G., T.D.F., and G.Z. constructed the scanning SOT microscope.

M.E.H. developed the SOT readout system.

A.V. and S.R. performed the TEM measurements

H.A., O.M., A.N., Y.A. and Y.P. wrote the paper with contributions from all authors.

# Supplementary material

## Interior and edge magnetization in thin exfoliated CrGeTe$_3$ films

*Avia Noah\*[1], Hen Alpern[1,2]\*\*, Sourabh Singh[1], Alon Gutfreund[1], Gilad Zisman[1], Tomer D. Feld[1], Atzmon Vakahi[3], Sergei Remennik[3], Yossi Paltiel[2], Martin Emile Huber[4], Victor Barrena[5], Hermann Suderow[5], Hadar Steinberg[1], Oded Millo[1], and Yonathan Anahory\*\*\*[1]*

[1]The Racah Institute of Physics, The Hebrew University, Jerusalem, 91904, Israel
[2]Department of Applied Physics, The Hebrew University of Jerusalem, Jerusalem 91904, Israel
[3]Center for Nanoscience and Nanotechnology, Hebrew University of Jerusalem, Jerusalem, 91904, Israel
[4]Departments of Physics and Electrical Engineering, University of Colorado Denver, Denver, CO 80217, USA
[5]Laboratorio de Bajas Temperaturas, Unidad Asociada UAM/CSIC, Departamento de Física de la Materia Condensada, Instituto Nicolás Cabrera and Condensed Matter Physics Center, Universidad Autónoma de Madrid, E-28049 Madrid, Spain

Corresponding authors: *avia.noah@mail.huji.ac.il, ** alprnhen@gmail.com,
*** yonathan.anahory@mail.huji.ac.il

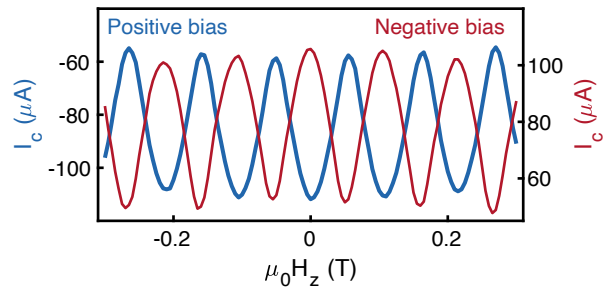

**Supplementary Figure S1. Quantum interference pattern of the SQUID-on-tip (SOT).** The critical current $I_c$ of one of the SOT's used in this work as a function of the applied out-of-Plane field $H_z$. Blue: Positive bias, red: Negative bias. The period of 105 mT of the quantum interference corresponds to an effective diameter of 155 nm of the SOT.

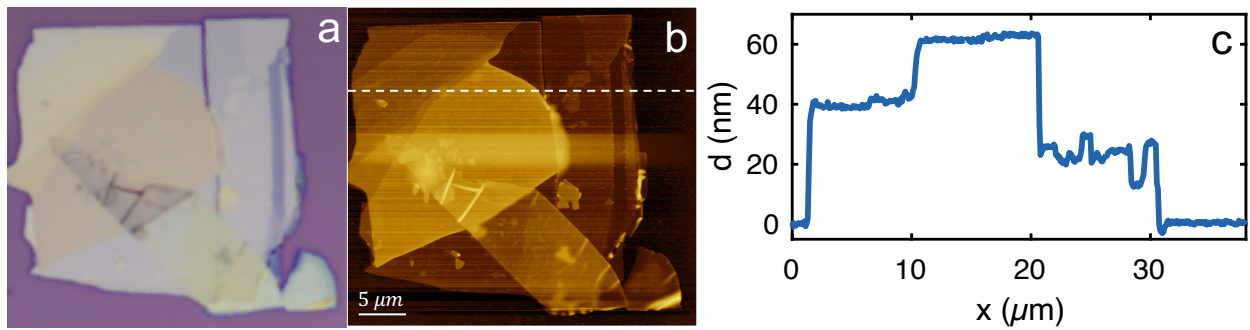

**Supplementary Figure S2.** AFM and optical thickness characterization. (a) Optical image of an exfoliated CGT flake on a SiO$_2$ substrate. The CGT flake contains varying thicknesses with correlated colors ranging from light purple (thinnest) to yellow (thickest). (b) A topography image of the flake seen in panel a, taken by an AFM using the tapping mode. (c) A topography cross section taken along the dashed line in panel b, sowing areas with distinct (uniform) thicknesses.

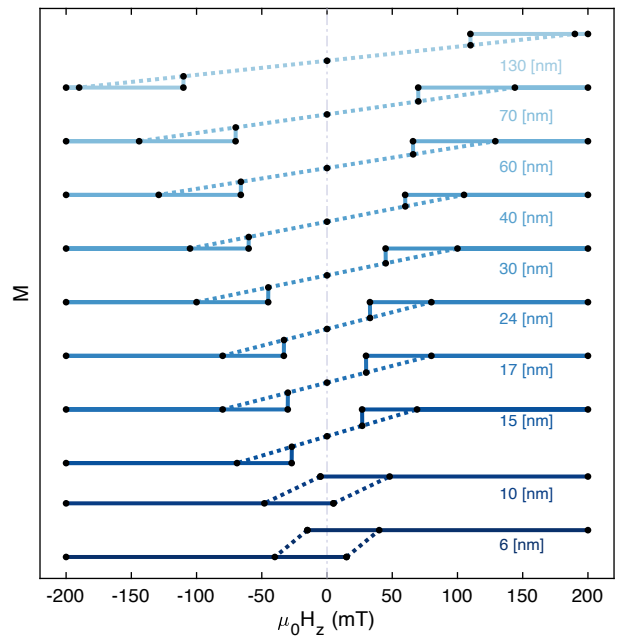

**Supplementary Figure S3. Transport of bilayer CGT/NbSe$_2$ at different thickness at 4.2 K.** Sketched magnetization curves drawn from $B_z(x,y)$ measured on film's parts of different $d$. Dashed lines are a guide to the eye connecting the two saturated fields. Thickness from bottom to top $d$ = 6, 10 ,15, 17, 24, 30, 40, 60, 70, 130 nm

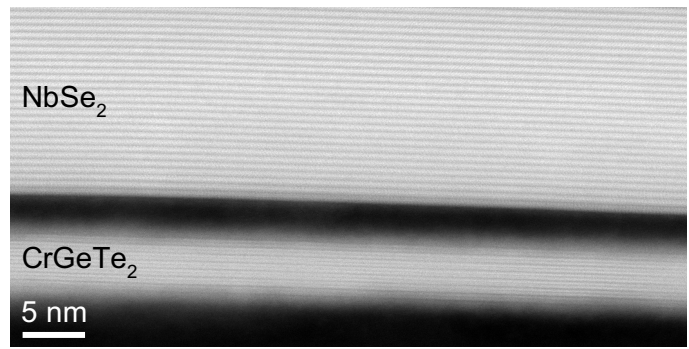

**Supplementary Figure S4.** High angle annular dark-HAADF image of the heterostructure in the sacrificial device where layers of CGT are seen to be separated by a gap from layers of NbSe$_2$.

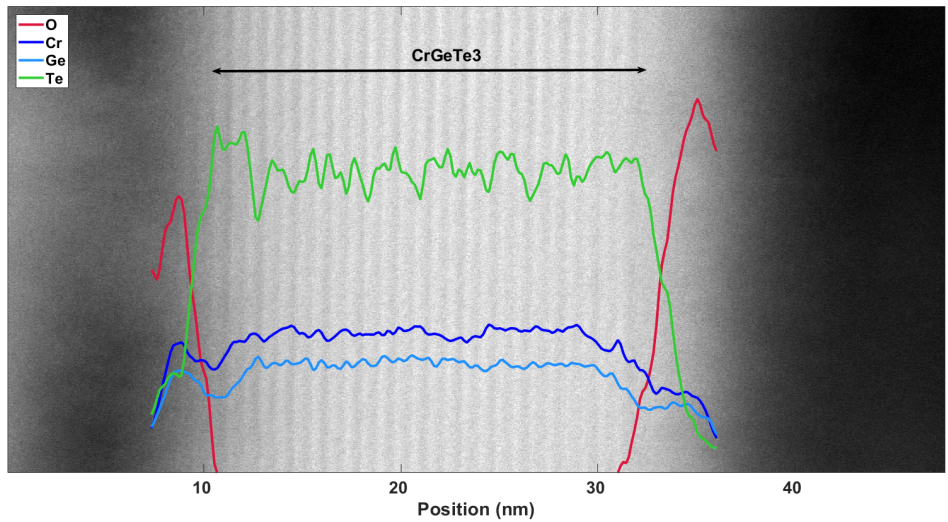

**Supplementary Figure S5.** EDS line scan, showing the relative amount of Cr, Ge, Te and O in a cross section of the device, in arbitrary units. The results are superimposed on a high-angle annular dark field (HAADF) image of the same cross section.

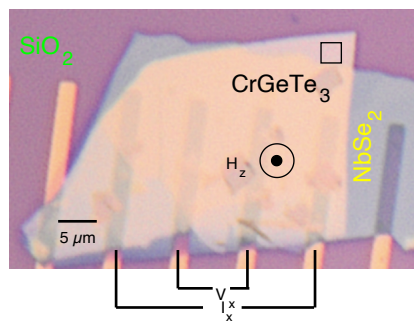

**Supplementary Figure 6.** Optical image of the bilayer and the Au electrodes used for the transport measurements presented in figure 1 and 2 of the main text. The black square corresponds to the region image in Fig. 1 b-g and Fig. 2. c-j of the main text.

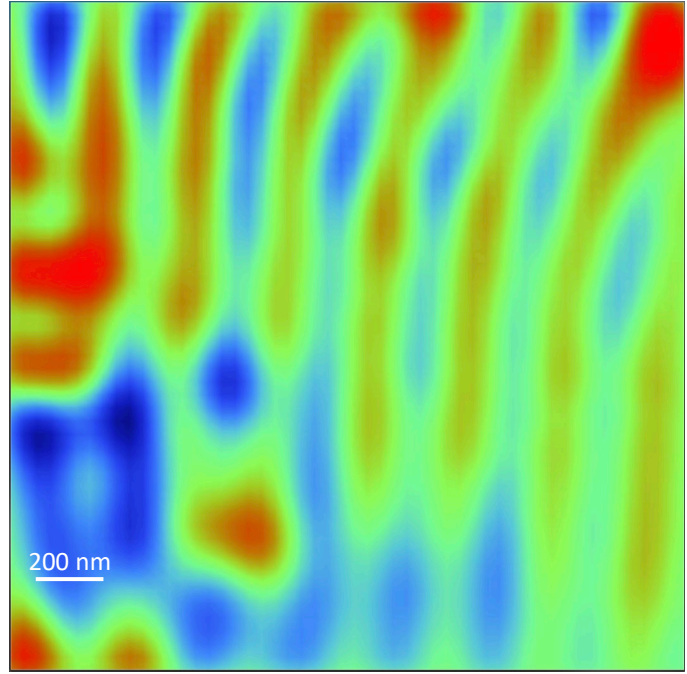

**Supplementary Figure S7. SQUID-on-tip image of 130 nm CrGeTe3 after zero field cooling (SOT).** Stripe magnetization with zero net magnetization in agreement with the theoretical prediction in the limit where the dipolar interaction exceeds the magnetic anisotropy. area scan 2x2 μm², pixel size 52 nm. The blue to red color scale represents lower and higher magnetic field respectively.

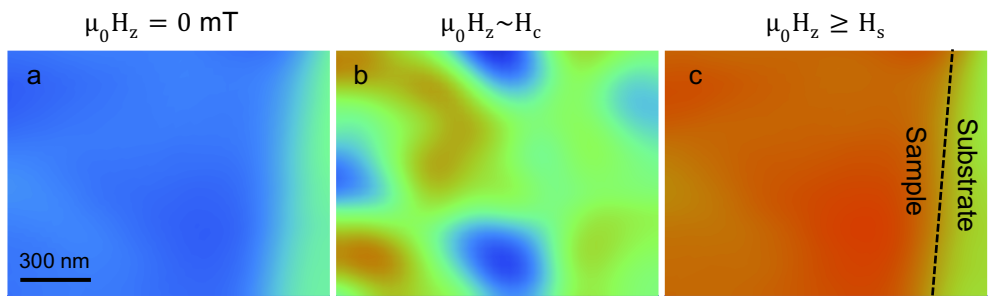

**Supplementary Figure S8. Scanning SOT microscopy images of a 6 nm thick CrGeTe₃ flake interior.** (**a-c**) Time sequence of magnetic images at distinct applied fields. (**a**) At zero field, after an excursion at $\mu_0 H_z = -200$ mT (larger than H$_s$). (**b**) Near the coercive field $H_c = 18$ mT (**c**) at 120 mT, above the saturation filed, where the sample is fully magnetized in the other direction. Imaging parameters: area scan 1.5×1 μm², pixel size 31 nm. The blue to red color scale represents lower and higher magnetic fields, respectively, while the green color represents the applied field. The full color scale is $B_z = 1$ mT in all images.

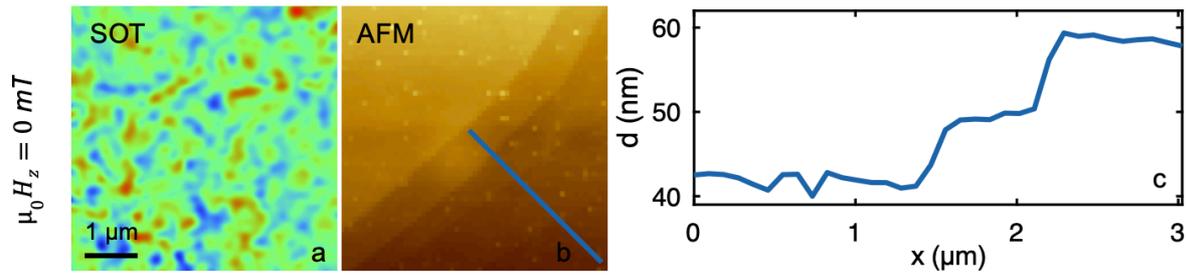

**Supplementary Figure S9. SOT image of CrGeTe₃ flake with a corresponding AFM image.** (a) SOT magnetic image at $\mu_0 H_z = 0$ mT. The magnetic domains evolve with no relation to the step-edges. (b) AFM image of the same scanning area resolving steps of ~ 10 nm. (c) Height profile measured along the blue line presented in **b**. Image parameters: area scan 5×5 µm², pixel size 26 nm. The blue to red color scale represents lower and higher magnetic field, respectively, with a scale of $B_z = 1$ mT.

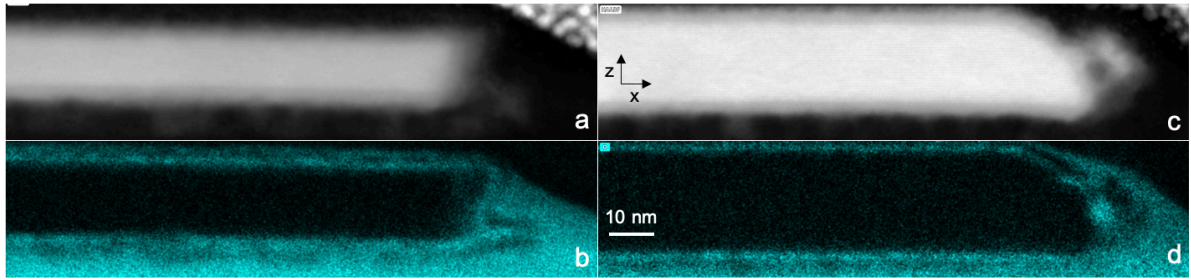

**Supplementary Figure S10. EDX and STEM of 17 and 24 nm flake edges.** (a-b) STEM cross-sectional images measured on the black lines presented in **Fig. 4 c,d**. (c-d) EDX measurements of the same cross-sections showing similar levels of oxidation on the edge and surface.

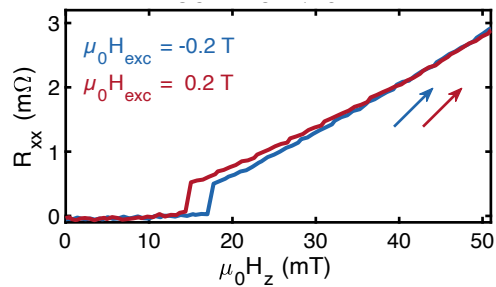

**Supplementary Figure S11.** $R_{xx}$ **magneto-transport measurements of bilayer CrGeTe₃/NbSe₂ showing magnetic memory at** $\mu_0 H_z = 0$ **for CGT of thickness** $d < 10$ **nm.** $R_{xx}$ as a function of the out-of-plane (OOP) field $H_z$ from zero to above the saturation field $H_s$ after different excursions at $\mu_0 H_{ex} = -0.2$ T and $\mu_0 H_{ex} = +0.2$ T, blue and red curves, respectively.

**Supplementary Note 1: SOT fabrication and characterization**

Supplementary figure 1 shows the quantum interference pattern typical for a SQUID. The SOT response to magnetic field is linked to the derivative of that pattern. To get good images, the SOT has to be field biased in a region where the interference pattern is linear, and one must avoid the regions where the response is zero (blind spots). The field period of this pattern is determined by the SQUID loop diameter so that each period represents a magnetic flux equal to the quantum of flux $\phi_0 = h/2e \approx 20.67$ G$\mu$m$^2$. This means that a smaller SQUID loop will yield a large field period with larger linear regions but with larger blind spots. Depending on the experimental requirements, optimal SQUID loop size is chosen. To mitigate the effect of blind spots, we use the fact that our SOTs are often composed of asymmetric junctions. This makes the interference pattern to have a different field offset for a different direction of the current running through the SQUID (Supplementary Figure 1, blue and red curves).

The magnetic length scale in CrGeTe$_3$ is on the order 100 nm or smaller. That implies a SQUID loop with a diameter below 100 nm. However, having a SQUID with a loop too small, of 50 nm for example, would yield a blind spot of +/- 0.4 T around zero field. That would be problematic to get good images below the saturation field (0.13 T). For this reason, all the SOTs used in this work have a SQUID loop of about 150 to 180 nm.

**Supplementary Note 2: Atomic force microscopy (AFM) measurements**

Topography measurements of CGT flakes exfoliated on Si-SiO$_2$ substrates were conducted to determine the thickness of the various areas measured by SOT. To that end a Ntegra modular apparatus (NT-MDT) was employed using the tapping method. Only areas showing thickness uniformity were used during the SOT characterization. An AFM (as well as an optical image) of one of the CGT flakes used in this work is shown in Supplementary Figure 2.

**Supplementary Note 3: STEM images of CGT with thickness analysis.**

To investigate the engagement between the NbSe$_2$ and CrGeTe$_3$ (CGT) we performed cross-section scanning transmission electron microscopy (STEM) and energy-dispersive spectroscopy (EDS) on the heterostructure device. High angle annular dark-field (HAADF) STEM image of a sacrificial device is shown in Supplementary Figure 4 which depicts a gap between the CGT and the NbSe$_2$. An average spacing of roughly 4 nm between the layers is seen in the figure. To understand the stoichiometry of the flakes EDS was performed and it reveals formation of an oxide layer on both sides of the flakes. Traces of Silicon and Carbon (not shown) were also observed in the EDS measurements which seem to originate because of the organic residue from the PDMS used during the exfoliation process.

To determine the thickness of thin CGT flakes we took the cross-sectional STEM images shown in Supplementary Figures 5 where individual layers of CGT can be seen. There is a formation of oxide layer on both sides of the flakes as seen from the fuzzy layers on top and bottom of the flake. The EDS line scan along the black arrow confirm the presence of Oxygen on the top and bottom of CGT flakes.

**Supplementary Note 4: Transport of a CGT/NbSe₂ bilayer with $d < 10$ nm**

We measured a CGT/NbSe₂ bilayer similar to the ones presented in Figures 1 and 2, but with a CGT layer below the critical thickness. The results are shown in Figure S11. First, we see a difference in $R_{xx}(H)$ when the sample was initialized at -0.2 T and +0.2 T. This clearly shows the presence of a global zero-field memory which is not observed for thicker samples (Figure 1 and 2). Second, we observe a raise in the field at which the voltage appears when the CGT was previously magnetized at $\mu_0 H_{ex} = -0.2$ T, opposite to the measurement field. This suggests that the sample was fully magnetized opposite to the field therefore reducing the field on the NbSe₂ and therefore the amount of vortices. Third, the two $R_{xx}(H)$ curves overlap above 35 mT, no matter what the previous history was. This suggests that the magnetic state of the CGT is fully polarized above that field which is consistent with the saturation field stated in Figure 3. Finally, around the coercive field, between ~20 to ~30 mT, the resistance after excursion at opposite field (-0.2 T, blue curve) is lower than that after excursion at 0.2 T (red curve), suggesting an enhancement of pinning due to the emergence of domains during the magnetization reversal process that takes place only when measuring the blue curve.

**Supplementary Movie 1:**

Movie of magnetic domains evolution imaged with the SOT and longitudinal resistance ($R_{xx}$) from zero field to the saturated state and back to zero in CrGeTe3. $\mu_0 H_z$ was ramped up from 0 to 130 mT (blue curve) and then, promptly, ramped down back to 0 (red curve). Yellow dot represents the field at which the image of the out-of-plane component of the magnetic field $B_z(x, y)$ was taken. Magnetic features of ~100 nm exhibiting a magnetic contrast (~5 mT) were observed. By increasing $H_z$, the domains that are anti-aligned with the field shrink while the ones that are parallel growing. From $\mu_0 H_z \sim 100$ mT, the magnetic contrast drops below 1 mT and the $B_z(x, y)$ stops evolving on that scale. By decreasing the field, the sample's magnetic images remain featureless down to $\mu_0 H_z = 40$ mT, where magnetic domains appear at once. A clear hysteresis is evident in the field range values between $\mu_0 H_z = 40$ mT and $\mu_0 H_z = 80$ mT, where a switching between the dissipationless and voltage states occurs. The frame size was $1 \times 1$ µm²; pixel 20 nm, acquisition time was 5 min/image and $H_z$ was increased in 5 mT step. The blue to red color scale represents lower and higher magnetic field, respectively; when high magnetic contrast appears, the color scale is intentionally saturated at the edge for clarity. Selected frames from the movie are shown Fig. 1 of the main text.

**Supplementary Movie 2:**

Movie of magnetic domains evolution imaged with the SOT and longitudinal resistance ($R_{xx}$) from zero field to the saturated state in 30 nm CrGeTe$_3$ after excursion of $\mu_0 H_{exc} = \pm 1$ T. $\mu_0 H_z$ was ramped up from 0 to 130 mT. The last dot represents the field at which the image of the out-of-plane component of the magnetic field $B_z(x, y)$ was taken. Magnetic features of ~ 100 nm exhibiting a magnetic contrast (~5 mT) were observed. By increasing $H_z$, the domains that are anti-aligned with the field shrink while the ones that are parallel growing. From $\mu_0 H_z \sim 100$ mT, the magnetic contrast drops below 1 mT and the $B_z(x, y)$ stops evolving on that scale. The same evolution both by images and transport were observed resolving zero magnetization at zero field. The frame size was $1 \times 1$ µm²; pixel 20 nm, acquisition time was 5 min/image and $H_z$ was increased in 5 mT step. The blue to red color scale represents lower and higher magnetic field, respectively; when high magnetic contrast appears, the color scale is intentionally saturated at the edge for clarity. Selected frames from the movie are shown Fig. 2 of the main text.

**Supplementary Movie 3:**

Movie of magnetic domains evolution imaged with the SOT between saturated states of 65 nm CrGeTe$_3$ thin film. $\mu_0 H_z$ was ramped up from 0 to 200 mT, then ramped down to $-200$ mT and back to zero. Magnetic features of ~ 100 nm were observed. By increasing $H_z$, the domains that are anti-aligned with the field shrink while the ones that are parallel growing. From $\mu_0 H_z \sim 145$ mT, the magnetic contrast drops below 1 mT and the $B_z(x, y)$ stops evolving on that scale. By decreasing the field, the sample's magnetic images remain featureless down to $\mu_0 H_z = 70$ mT, where magnetic domains appear at once. The same behavior was observed for the negative sweep. The frame size was $5 \times 5$ µm²; pixel 40 nm, acquisition time was 5 min/image and $H_z$ was increased in 5 mT step. The blue to red color scale represents lower and higher magnetic field, respectively; when high magnetic contrast appears, the color scale is intentionally saturated at the edge for clarity. Selected frames from the movie are shown Fig. 3a-d of the main text.

**Supplementary Movie 4:**

Movie of magnetic domains evolution imaged with the SOT between saturated states of 25 nm CrGeTe$_3$ thin film. $\mu_0 H_z$ was ramped up from 0 to 120 mT, then ramped down to $-120$ mT and back to zero. Magnetic features of ~ 100 nm were observed. By increasing $H_z$, the domains that are anti-aligned with the field shrink while the ones that are parallel growing. From $\mu_0 H_z$~ 80 mT, the magnetic contrast drops below 1 mT and the $B_z(x,y)$ stops evolving on that scale. By decreasing the field, the sample's magnetic images remain featureless down to $\mu_0 H_z = 30$ mT, where magnetic domains appear at once. The same behavior was observed for the negative sweep. The frame size was $2 \times 2$ µm²; pixel 30 nm, acquisition time was 5 min/image and $H_z$ was increased in 5 mT step. The blue to red color scale represents lower and higher magnetic field, respectively; when high magnetic contrast appears, the color scale is intentionally saturated at the edge for clarity. Selected frames from the movie are shown Fig. 3e-h of the main text.

**Supplementary Movie 5:**

Movie of magnetic domains evolution imaged with the SOT between saturated states through an intermediate domain phase in 6 nm CrGeTe$_3$ thin film. $\mu_0 H_z$ was ramped up from 0 to 120 mT, then ramped down to $-120$ mT and back to zero. At zero field no magnetic features were observed. At $\mu_0 H_z = 20$ mT magnetic domains appear. By increasing the field domains that are anti-aligned with the field shrink while the ones that are parallel growing. From $\mu_0 H_z$~ 45 mT, the magnetic contrast drops below 1 mT and the $B_z(x,y)$ stops evolving on that scale. By decreasing the field to zero no magnetic features were observed resolving ferromagnetic behavior. The same behavior was observed for the negative sweep. The frame size was $1 \times 1$ µm²; pixel 30 nm, acquisition time was 5 min/image and $H_z$ was increased in 5 mT step. The blue to red color scale represents lower and higher magnetic field, respectively; when high magnetic contrast appears, the color scale is intentionally saturated at the edge for clarity. Selected frames from the movie are shown Fig. 3i-l of the main text.